\documentclass[conference]{IEEEtran}
\IEEEoverridecommandlockouts
\usepackage{cite}
\usepackage{amsmath,amssymb,amsfonts}
\usepackage{graphicx}
\usepackage{textcomp}
\usepackage{xcolor}
\usepackage{algorithm}
\usepackage{algpseudocode}
\usepackage{amsmath}
\usepackage{algorithmicx}
\usepackage{graphicx}
\usepackage{longtable}
\usepackage{geometry}
\usepackage{pgfplots}
\pgfplotsset{compat=newest}
\usepackage{parskip}
\usepackage{verbatim}
\usepackage{pdflscape}
\usepackage{tabularx}
\usepackage{ltablex}
\usepackage{array}
\usepackage{enumitem}
\usepackage{multirow}
\usepackage{adjustbox}
\usepackage{lipsum}
\usepackage{pgfplots}
\usepackage{pgfplotstable}
\usepgfplotslibrary{groupplots}
\usepackage{tikz}
\usetikzlibrary{shapes.geometric, arrows, patterns, shapes.arrows, positioning}
\tikzstyle{startstop} = [rectangle, rounded corners, minimum width=3cm, minimum height=1cm,text centered, draw=black, fill=red!30]
\tikzstyle{process} = [rectangle, minimum width=3cm, minimum height=1cm, text centered, draw=black, fill=blue!30]
\tikzstyle{decision} = [diamond, minimum width=3cm, minimum height=1cm, text centered, draw=black, fill=green!30]
\tikzstyle{arrow} = [thick,->,>=stealth]
\def\BibTeX{{\rm B\kern-.05em{\sc i\kern-.025em b}\kern-.08em
    T\kern-.1667em\lower.7ex\hbox{E}\kern-.125emX}}

\newcommand{\ieeedoi}{10.1109/BRAINS67003.2025.11302948}
\newcommand{\ieeecitation}{D. Mancino, H. O. Sevim, and O. Saguillo Gonzalez,
``Bunny Hops and Blockchain Stops: Cross-Chain MEV Detection With N-Hops,''
in \emph{Proceedings of the 2025 7th Conference on Blockchain Research \& Applications
for Innovative Networks and Services (BRAINS)}, 2025, pp.~1--4, doi: \ieeedoi.}
\newcommand{\ieeenotice}{\copyright\ 2025 IEEE. Personal use of this material is permitted.
Permission from IEEE must be obtained for all other uses, in any current or future media,
including reprinting/republishing this material for advertising or promotional purposes,
creating new collective works, for resale or redistribution to servers or lists,
or reuse of any copyrighted component of this work in other works.}

\begin{document}

\title{Bunny Hops and Blockchain Stops: Cross-Chain MEV Detection With N-Hops%
\thanks{\ieeenotice\ \ \ieeecitation}}

\author{
    Davide Mancino\textsuperscript{1,2},
    Hasret Ozan Sevim\textsuperscript{2,3},
    Oriol Saguillo Gonzalez\textsuperscript{4} \\
    \textsuperscript{1}\textit{University of Milano-Bicocca, DISCo, Milano, Italy} \\
    \textsuperscript{2}\textit{University of Camerino, Camerino, Italy} \\
    \textsuperscript{3}\textit{Catholic University of Sacred Heart, Milano, Italy} \\
    \textsuperscript{4}\textit{IMDEA Networks Institute, Madrid, Spain} \\
    davide.mancino@unimib.it, 
    hasretozan.sevim@unicam.it, 
    oriol.saguillo@imdea.org
}


\maketitle

\begin{abstract}
This student paper introduces a novel methodology for the detection and analysis of multihop cross-chain arbitrage opportunities, wherein multihop denotes arbitrage sequences involving more than two transactional steps across distinct blockchain networks, executed using sequence-dependent strategies. Utilizing a comprehensive dataset comprising over 2.4 billion transactions recorded between September 2023 and August 2024 (encompassing 12 blockchain platforms and 45 cross-chain bridges) we design and implement an algorithm capable of identifying, sequence-dependent arbitrage paths spanning multiple ecosystems. Our empirical analysis demonstrates that such arbitrage opportunities are exceedingly infrequent, underscoring the inherent challenges associated with multihop execution in cross-chain environments.

\end{abstract}

\begin{IEEEkeywords}
Decentralized Finance, Maximal Extractable Value, Arbitrage, Cross-Chain
\end{IEEEkeywords}

\section{Introduction}

A blockchain can be viewed as a decentralized database, where any user holding the native token can initiate transactions or interact with applications deployed on the network. The emergence of various types of blockchains reflects efforts to target specific markets or industry use cases. However, this growing diversity introduces a major challenge: ensuring interoperability across these separate ecosystems. Recent advancements in cross-chain bridge technologies (such as LayerZero and others) have significantly improved the user experience by enabling more seamless asset transfers and communication between blockchains.

Maximal extractable value (MEV) has been a critical issue in the security of blockchain systems, to protect the liquidity providers in different decentralized applications. The surge of the multi-blockchain ecosystem is creating new liquidity connections. Cross-chain MEV has emerged as a new field of research, driven by the interoperability of Ethereum's Layer 2 networks and other blockchains. Recent methodologies have been developed to detect, analyze, and optimize different opportunities, but so far these methods focus on two-hop arbitrages. 

Our contributions in this work are as follows: (1) Presenting the fundamental knowledge and the related work that we build upon, (2) Explaining the data and diving into the Sequence-Dependent Arbitrage (SDA) multihop model, (3) Discussing and evaluating the detected SDA multihop arbitrages, and (4) Validating the significance of cross-chain SDA multihop arbitrage in the cross-chain MEV field and outlining future research directions.

\section{Background and Related Work}

This section will provide relevant background about Ethereum Layer 2, MEV, liquidity fragmentation, and the types of sequence cross-chain arbitrage.

\subsection{Ethereum Layer 2}

Ethereum \cite{wood2014ethereum} is currently the second-largest blockchain by market capitalization and supports the development of smart contracts. Smart contracts enable decentralized applications (dApps), which run not on a single server but across a network of computers connected via a peer-to-peer architecture. In order to overcome the scalability issues of Ethereum, rollups (also known as Layer 2 solutions) have emerged as the primary scaling approach for Ethereum. A Layer 2 can be either a general-purpose blockchain or tailored for a specific application. General-purpose Layer 2s include optimistic rollups, such as Arbitrum,
and zero-knowledge (ZK) rollups, such as Starknet. In contrast, application-specific Layer 2s resemble the design philosophy of the Cosmos ecosystem, with examples like dYdX, which operates its own custom chain optimized for a particular use case. 

The connectivity between Ethereum and its Layer 2 solutions is typically established through bridges, which are smart contracts deployed on Ethereum (Layer 1) that enable the transfer of assets and data between L1 and L2. In general, users can deposit funds into a bridge contract on Ethereum, after which they can interact with the Layer 2 network. Transactions are processed on Layer 2, and the resulting state changes are periodically submitted back to Ethereum for final settlement.

\subsection{Maximal Extractable Value}
Maximal Extractable Value (MEV) \cite{daian2020flash} denotes the profit obtained by reordering, inserting, or censoring transactions within a block. It arises when validators or other privileged actors manipulate transaction order for financial gain~\cite{mancino:role_reward_mev:dlt:2024}. MEV practices are often classified as either malicious or beneficial. Malicious forms, such as frontrunning and sandwich attacks, extract value from users and liquidity providers. Beneficial MEV, like arbitrage, can improve market efficiency by correcting price discrepancies between decentralized and centralized exchanges \cite{heimbach2024non}.

\subsection{Liquidity Fragmentation}

The augmentation of Automated Market Makers (AMMs) on multiple blockchains has raised concerns about the issue known as liquidity fragmentation. As Liquidity Providers (LPs) become increasingly dispersed across chains, price discrepancies among AMMs can grow, leading to problems such as loss-versus-rebalancing~\cite{milionis2024automated}. This challenge is further intensified by the emergence of additional Layer 2 networks, which introduce increased competition among platforms to attract and retain LPs.

\subsection{Types of Sequence Cross-chain Arbitrage}

\"Oz et al.~\cite{oz2025pandorasboxcrosschainarbitrages} distinguish two forms of cross-chain arbitrage: Sequence-Independent Arbitrage (SIA) and Sequence-Dependent Arbitrage (SDA). In SIA, the arbitrageur holds inventory on both chains, allowing the two legs of the trade to be executed in any order. This decoupling enables flexibility but requires periodic rebalancing of assets. In contrast, SDA involves transferring assets across chains—typically via bridges—so that the output of the hedge leg funds the profit leg. Since SDA relies on ordered, cross-chain execution, it is inherently sequential and more exposed to latency and risk. This paper focuses on SDA to analyze these properties in depth.

\section{Data}

We analyze cross-chain arbitrage opportunities using Flipside Crypto\footnote{https://flipsidecrypto.xyz/} data spanning September 1, 2023 to August 31, 2024, which corresponds to the same time range adopted in study~\cite{oz2025pandorasboxcrosschainarbitrages} on the detection of two-hop arbitrage strategies.
The dataset comprises 2,490,877,965 unique swaps and 34,835,278 bridge transactions across 12 blockchain networks: Ethereum, Arbitrum, Near, Polygon, BSC, Solana, Blast, Osmosis, Avalanche, Optimism, Base, and Gnosis. Our analysis covers 45 different bridge protocols including Symbiosis, Wormhole, Axelar, Stargate, Multichain, and LayerZero. This comprehensive coverage enables rigorous examination of cross-chain arbitrage patterns and inefficiencies across the fragmented DeFi ecosystem.

\section{Multihop Model}
\label{sec:multihopmodel}

We assume that a set of sophisticated actors will attempt to identify optimal arbitrage paths in a cross-chain environment, this problem has been shown to be computationally feasible~\cite{angeris2022optimal}. This environment involves a significantly larger number of AMMs compared to a single-chain scenario, increasing the complexity of path discovery.
Our central hypothesis builds upon the findings of \cite{oz2025pandorasboxcrosschainarbitrages}, which estimate that Sequence-Dependent Arbitrage (SDA) accounts for approximately 30\% of two-hop cross-chain arbitrage opportunities. However, when considering multihop arbitrage, SDA faces considerable challenges compared to Sequence-Independent Arbitrage (SIA). These challenges stem primarily from the higher costs associated with bridge fees and the latency involved in waiting for profitable opportunities to arise.

We propose that the feasibility of SDA in multihop scenarios follows a power-law distribution: as the number of required hops increases, the probability of successfully executing the arbitrage decreases significantly. This reflects the growing cost, complexity, and risk associated with longer cross-chain arbitrage paths.

\section{Identifying Sequence Dependent Arbitrage Multihop} \label{sec:identSDAarbitrage}

\subsection{Identifying Multihop Trades}
We define the SDA path as on-chain swap and bridge transactions executed by a single actor on multiple blockchains. A general $n$-hop arbitrage path consists of $2n - 1$ transactions alternating between swaps and bridges:

\begin{equation*}
A = \{t_1, t_2, t_3, ..., t_{2n-1}\}
\end{equation*}

where $t_{2i-1}$ denotes a swap transaction on chain $C_i$ and $t_{2i}$ denotes a bridge transaction from $C_i$ to $C_{i+1}$,
 for all $i = 1, ..., n$.

Each transaction $t_i$ is represented as a tuple:
\[
t_i = (h_i, \tau_i, S_i, R_i, C_i, X_{in}^{(i)}, X_{out}^{(i)}, v_{in}^{(i)}, v_{out}^{(i)}, \alpha_i, \beta_i)
\]

where $h_i$ is the transaction hash, $\tau_i$ is the timestamp, $S_i$ is the sending address, $R_i$ is the receiving address, $C_i$ is the blockchain on which $t_i$ occurred, $X_{in}^{(i)}$ and $X_{out}^{(i)}$ are the input and output tokens, $v_{in}^{(i)}$ and $v_{out}^{(i)}$ are the corresponding input and output values (in USD), $\alpha_i$ is a binary indicator equal to 1 if the transaction is a swap and 0 otherwise, and $\beta_i$ is a binary indicator equal to 1 if the transaction is a bridge and 0 otherwise.
We impose the constraint that $\alpha_i + \beta_i = 1$ (each transaction must be either a swap or a bridge, but not both).
To ensure interpretability and reduce false positives, we impose two heuristics on admissible arbitrage paths:

\begin{itemize}
    \item Only atomic two-token swaps are considered. Transactions involving more than two assets (e.g., multi-path swaps or aggregators) are excluded.
    \item The sequence must involve effective token changes. If $X_{in}^{(2i-1)} = X_{out}^{(2i-1)}$ for all swap steps, the path is discarded, as it does not reflect economic arbitrage.
\end{itemize}

\subsection{Transaction Sequence Constraints}
For any two consecutive transactions $(t_i, t_{i+1})$, we enforce multiple constraints to ensure the sequence represents a legitimate arbitrage:
\begin{description}[style=unboxed,leftmargin=0cm]

\item[Constraint 1 – Time Continuity:] 
\begin{align}
\tau_i < \tau_{i+1} &\leq \tau_i + \Delta t_i
\end{align}
We require that transactions follow temporal order and fall within a conservative window of $\Delta t_i = 5$ minutes, chosen primarily to establish the existence of cross-chain n-hop sequences while minimizing false positives in our detection.
\item[Constraint 2 – Value Continuity:] 
\begin{align}
 v_{in}(t_{i+1}) \in [0.98 \cdot v_{out}(t_i), v_{out}(t_i)]
\end{align}
Accounts for a value discrepancy of up to 2\%, which may arise from transaction fees or intentional approximations by the sender during execution.

\item[Constraint 3 – Token Continuity:] 
\begin{align}
X_{out}(t_i) &= X_{in}(t_{i+1})
\end{align}
Ensures that tokens flow consistently between consecutive steps.

\item[Constraint 4 – Actor Consistency:] 
\begin{equation}
S_{i+1} =
\begin{cases}
S_i \quad \text{if } \alpha_i = 1 \text{ and } \beta_{i+1} = 1 \\
R_i \quad \text{if } \beta_i = 1 \text{ and } \alpha_{i+1} = 1
\end{cases}
\end{equation}
This ensures that the sender of transaction $t_{i+1}$ is either the same as the sender of $t_i$ (when $t_i$ is a swap followed by a bridge), or the receiver of $t_i$ (when $t_i$ is a bridge followed by a swap).

\item[Constraint 5 – Cross-Chain Consistency:]
\begin{equation}
C_{i+1} =
\begin{cases}
C_i \quad \text{if } \alpha_i = 1 \text{ and } \beta_{i+1} = 1 \\
\neq C_i \quad \text{if } \beta_i = 1 \text{ and } \alpha_{i+1} = 1
\end{cases}
\end{equation}
When a swap is followed by a bridge, both must occur on the same chain. When a bridge is followed by a swap, they must occur on different chains.

\end{description}

\subsection{Validation Function}
We define an indicator function $\phi(t_i, t_{i+1})$ that evaluates whether two consecutive transactions satisfy constraints (1)-(5), from the whole set of transactions:
\begin{equation*}
\phi(t_i, t_{i+1}) =
\begin{cases}
1 \quad \text{if constraints (1)--(5) are satisfied} \\
0 \quad \text{otherwise}
\end{cases}
\end{equation*}
The global flow function over the arbitrage path $A = {t_1, ..., t_{2n-1}}$ becomes:
\begin{equation*}
\Phi(A) = \prod_{i=1}^{2n-2} \phi(t_i, t_{i+1})
\end{equation*}
An arbitrage is valid if $\Phi(A) = 1$, meaning all consecutive transactions satisfy our constraints.

\subsection{Profit}
\textbf{Gross profit:}
\[
\Pi(A) = v_{out}(t_{2n-1}) - v_{in}(t_1)
\]
This represents the absolute profit gained from the arbitrage sequence, calculated as the difference between the final output value and the initial input value.

\section{Results}

\begin{figure}[htbp]
\centering
\includegraphics[width=0.9\linewidth]{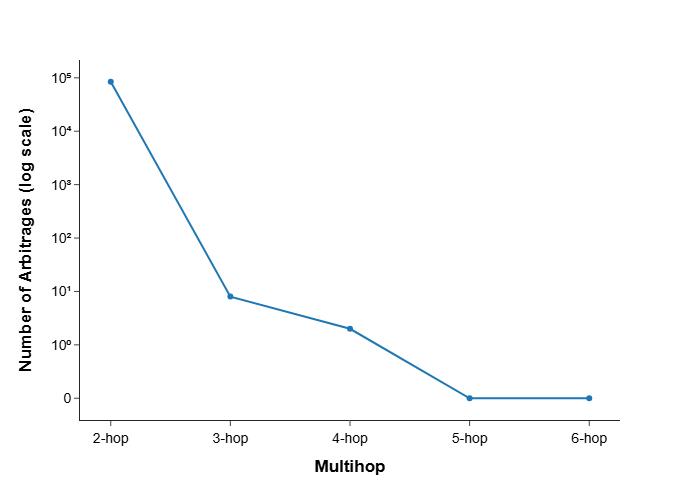}
\caption{Log-scale distribution of identified cross-chain arbitrages as a function of multihop path complexity. The count for 2-hop SDAs is derived from the data presented in Öz et al.~\cite{oz2025pandorasboxcrosschainarbitrages}.}
\label{fig:multihoppowelaw}
\end{figure}

Our analysis of cross-chain arbitrage patterns spanning September 2023 to August 2024 revealed important insights into multihop arbitrage behavior across the fragmented DeFi ecosystem. Using the methodology described in Section \ref{sec:identSDAarbitrage}, we identified a total of 10 multihop cross-chain arbitrage instances: 8 three-hop and 2 four-hop transactions. The investigation found no instances of five-hop or six-hop arbitrages, suggesting that increasing complexity negatively impacts the feasibility of cross-chain arbitrage strategies. As discussed in Section~\ref{sec:multihopmodel}, the number of identified arbitrages follows a power-law distribution with respect to hop count, as confirmed by our statistical analysis (AIC = 4.76, RMSE = 1.13 compared to AIC = 4.86, RMSE = 1.15 for exponential model), while execution time increases approximately linearly with the number of hops, making a profit becomes increasingly difficult due to the latency required for the full path to complete (Fig.~\ref{fig:multihoppowelaw}).

The 8 identified three-hop arbitrages each show distinct patterns in their token combinations, blockchain pathways, and profitability, as detailed in Table~\ref{tab:3hop_arbitrage}.

\begin{table}[ht]
\caption{Summary of Three-Hop Arbitrage Transactions}
\label{tab:3hop_arbitrage}
\centering
\scriptsize
\begin{tabular}{@{}p{2.3cm}cp{2.5cm}@{\hspace{0.5em}}c@{}}
\hline
\textbf{Chain Path} & \textbf{Dur.(s)} & \textbf{Tokens} & \textbf{Profit(\$)} \\
\hline
Base$\to$Eth$\to$Base & 646 & USDC/BAL & 32.78 \\
Base$\to$Opt$\to$Base & 490 & HOP/WETH & 0.43 \\
Arb$\to$Opt$\to$Base & 311 & ARB/WETH/HOP & -255.07 \\
Opt$\to$Base$\to$Arb & 466 & WETH/HOP & 264.04 \\
Base$\to$Avax$\to$Base & 448 & USDC/axlUSDC & 21.40 \\
Arb$\to$Poly$\to$Arb & 412 & USDT/USDC/WBTC & -0.17 \\
Blast$\to$Arb$\to$Eth & 458 & WETH/ezETH & -8.17 \\
Poly$\to$Arb$\to$Poly & 242 & WMATIC/WBTC/WETH & -0.02 \\
\hline
\end{tabular}
\end{table}

Base was the most frequently used blockchain, appearing in 5 out of 8 three-hop instances, followed by Arbitrum (4 instances) and Optimism (3 instances). The mean duration in three-hop arbitrages was 434.1 seconds (approximately 7.2 minutes), highlighting the extended execution time compared to single-chain arbitrages. This extended duration is primarily attributable to cross-chain bridge latency and the sequential nature of transactions.

The two identified four-hop arbitrages displayed greater complexity in the path and duration. The first four-hop transactions routed Optimism $\rightarrow$ Base $\rightarrow$ Optimism $\rightarrow$ Base with a duration of 776 seconds (12.9 minutes), yielding a profit of \$20.69. The second instance traveled Arbitrum $\rightarrow$ Optimism $\rightarrow$ Base $\rightarrow$ Arbitrum in 617 seconds (10.3 minutes) with a profit of \$1.61. The same address executed both four-hop transactions.
(0xf6c77e...ace28d), suggesting that he is a sophisticated actor.

Profitability analysis revealed that 6 of 10 multihop arbitrages generated positive returns, with profits ranging from \$0.43 to \$264.04. The highest profit (\$264.04) was achieved through a three-hop arbitrage involving Optimism, Base, and Arbitrum using WETH and HOP tokens. On the other hand, the largest loss (\$-255.07) occurred in a three-hop arbitrage involving Arbitrum, Optimism, and Base with ARB, WETH, and HOP tokens.
The observed patterns indicate that cross-chain arbitrage becomes progressively more challenging as the number of hops increases, primarily due to accumulated transaction costs, greater execution risk, and longer completion times, all of which amplify exposure to market volatility. These findings align with the observations made by \"{O}z et al.~\cite{oz2025pandorasboxcrosschainarbitrages}, who noted similar challenges in two-hop cross-chain arbitrages.

\section{Conclusion}
\label{sec:conclusion}
Our research presents the first systematic analysis of sequence-dependent, multihop cross-chain arbitrages throughout the fragmented DeFi ecosystem. By examining more than 2.4 billion swaps and 34.8 million bridge transactions spanning 12 blockchain networks, we identified only 10 multihop arbitrage instances (8 three-hop and 2 four-hop), demonstrating their exceptional rarity.

The scarcity of multihop arbitrages stems from: (1) accumulated transaction costs, (2) extended execution times (averaging 7.2 minutes for three-hop and 11.6 minutes for four-hop transactions), (3) increased exposure to price volatility, and (4) heightened technical complexity. These challenges were reflected in profitability metrics, with only 60\% of identified transactions yielding positive returns.

Future research should: (1) extend the methodology to include sequence-independent arbitrages (SIAs) for more comprehensive analysis, and (2) incorporate bridge latency and fee structures in profitability calculations. As blockchain interoperability improves, we anticipate enhanced feasibility of complex cross-chain strategies.

\bibliographystyle{IEEEtran}
\bibliography{bibliography}

\end{document}